\documentstyle[preprint,aps]{revtex}

\input epsf

\tightenlines
\begin{document}
\draft
\title{Electromagnetic properties of hadrons via the $u-d$ mass difference
       and direct photon exchange.}
\author{R.~Delbourgo\cite{Author1}, Dongsheng~Liu\cite{Author2} and 
        M.D.~Scadron\cite{Author3}}
\address{University of Tasmania, GPO Box 252-21, Hobart,\\
Tasmania 7001, Australia}
\date{\today }
\maketitle

\begin{abstract}
We demonstrate that a $u-d$ quark mass difference (which we 
estimate to be about 4 MeV) within quantum loops, can reproduce 
the effects of the Coleman-Glashow electromagnetic tadpole 
operator.
\end{abstract}


\narrowtext

\section{INTRODUCTION}


It is now well understood that the neutron is heavier than the proton
because the down quark is heavier than the up quark, which more than
compensates the positive electromagnetic charge energy of the proton. 
An earlier, alternative interpretation of the idea is due to Coleman 
and Glashow\cite{CG} who express this as an electromagnetic 
Hamiltonian operator,
\begin{equation}
 H_{em} = H_{JJ} + H^3_{tad},
\end{equation}
which contains an intrinsic `tadpole' term transforming as the third
component of isospin ($\lambda_3$ in an SU(3) context); the $H_{JJ}$ 
piece corresponds to the first order in $\alpha$ contribution 
due to photon exchange, which can be estimated in various ways and is
predominantly positive. By using the group theoretical properties of
$H_{tad}^3$ one can thereby correlate many mass differences and magnetic
moments of hadrons within a multiplet.

In this paper we shall try to ascribe the tadpole term to the $u-d$
constituent quark mass difference and therefore obtain a more dynamical
picture of its origin. This will involve us in evaluating {\em differences}
of certain `bubble graphs' plus vacuum `tadpole' diagrams
for hadron self-energies, in addition to virtual photon emission/absorption
graphs. Apart from the vacuum diagrams, these bubble graph differences
contain logarithmic infinities, requiring regularization. (The results
are actually insensitive to the regularization and do not depend on
momentum routing.) In any case, as needed, we will at various points
make use of simple quark chiral symmetry ideas\cite{DS1}, such as the 
Goldberger-Treiman relation (for $f_\pi \simeq 93$ MeV and 
$g=2\pi/\sqrt{3}$),
\begin{equation}
 f_{\pi}g = (m_u+m_d)/2 \equiv \hat{m} \simeq 337 {\rm ~MeV},
\end{equation}
that correspond to `gap equations' in a dynamical context, in order to 
handle logarithmic infinities which sometimes arise. (The magnitude
(2) also arises in the context of magnetic moments\cite{DeRu}.) We will 
see that this quark interpretation of the Coleman-Glashow tadpole works 
quite well and provides a more fundamental description of electromagnetic properties of hadrons.

However, before starting, it is as well to remind ourselves about the
sort of magnitudes that we are chasing. The $p-n$ mass difference of
about -1.3 MeV gives a first indication: since virtual photon exchange
provides the positive terms
\begin{equation}
 \langle p|H_{JJ}|p \rangle \simeq 1.2 {\rm~MeV}, \quad
 \langle n|H_{JJ}|n \rangle \simeq 0,
\end{equation}
we may deduce\cite{CS}
\begin{equation}
 \langle p|H_{tad}^3|p\rangle-\langle n|H_{tad}^3|n\rangle\simeq 
 -2.5{\rm~MeV}.
\end{equation}
The naive conclusion is therefore that $m_d - m_u \sim 3$ MeV. In the same
vein, the observed $\Sigma^- - \Sigma^+$ mass splitting of 8 MeV suggests
that
\begin{equation}
 m_d - m_u \simeq (m_{\Sigma^-}-m_{\Sigma^+})/2 \sim 4 {\rm~MeV},
\end{equation}
since the $H_{JJ}$ contribution, due alone to magnetic moments, is small; 
it also accords with the SU(3) symmetry prediction\cite{f0}, 
$(H^3_{tad})_{\Sigma^+-\Sigma^-}\simeq 3(H^3_{tad})_{p-n}\simeq -7.5{\rm~MeV}$.
Finally, we may extract a crude value for $m_d-m_u$ from the observed
$K^0 - K^+$ mass difference, since both $d\bar{s}$ and $u\bar{s}$ have the
same quark structure; thus
$$m_d - m_u \simeq m_{K^0} - m_{K^+} \simeq 4 {\rm~MeV}.$$

We will review expectations for the $u-d$ mass difference in the next 
section. Then in section III we shall study pseudoscalar meson 
electromagnetic mass splittings from the perspective of QED and quark 
loop diagrams. A similar approach is applied to vector mesons in 
section IV, with particular emphasis on $\Delta I=1,\,\omega-\rho^0$ 
mixing. We will find that all cases, associated with logarithmic
divergences, are governed by a universal (tadpole) scale of about
-5200 MeV$^2$, which is determined by a $d-u$ mass difference of about
4 MeV. There we also offer some comparisons with other group-theoretical
approaches.

\section{Constituent quark $d-u$ mass difference}


The nonstrange and strange constituent quark masses are very roughly given by
$\hat{m} \simeq m_\rho/2 \sim 380$ MeV, $m_s \simeq m_\phi/2 \sim 510$
MeV, respectively. Similar results follow by examining the baryon 
masses but more accurate scales are found from the Goldberger-Treiman 
relation (2) and its analogue, $f_Kg = (m_s + \hat{m})/2$. Since 
experiments\cite{PDG} give $f_K/f_\pi\simeq 1.22$, one deduces that 
$m_s/\hat{m} \simeq 1.44$, leading to the mass scales (within a few MeV),
$\hat{m} \simeq 340 {\rm~MeV},\,m_s \simeq 490$ MeV.

To obtain the still smaller $u-d$ mass difference, it is necessary to look
at em mass differences between baryons, as indicated above, which suggest
that $m_d - m_u \sim 4$ MeV. A similar inference can be drawn from the
pseudoscalar mass differences: firstly we note that $H_{tad}$ does not
contribute to the $\pi^+-\pi^0$ mass difference; secondly we take it that
the $H_{JJ}$ term gives a comparable magnitude both to $\pi^+-\pi^0$ and
$K^+ - K^0$ squared masses. It follows that the tadpole contribution 
to $m^2_{K^+} - m^2_{\pi^+}$ is determined by\cite{CS}
$\Delta m_K^2 - \Delta m_\pi^2 \simeq -5200$ MeV$^2$. Because we are
ascribing this part to the underlying quarks, we may roughly equate it to
(as we shall confirm in different guise in the next section) 
$2(m_u-m_d)m_K$, whereupon one 
deduces $m_d - m_u \sim 4$ MeV. Dashen's electromagnetic (em) PCAC 
theorems\cite{RD} are in conformity with this result.

Another source of information is the amazingly accurate hyperfine
splitting quark model\cite{Is} which predicts the constituent mass difference
$m_d - m_u \simeq 6$ MeV, but in conjunction with a (baryon) quark mass scale
of $\hat{m} \simeq 363$ MeV. Since this is rather greater than the previous scale
by a factor of about 10\%, one is inclined to reduce the Isgur value of
$m_d - m_u$ to 5.5 MeV or less. The Particle Data Tables\cite{PDG}
provide yet another source, but for
the current quark mass difference; they say that $(m_d-m_u)_{current}$
hovers around 5 MeV with an error of about 2 MeV. A more global approach
due to Lichtenberg\cite{Li} finds that the constituent $d-u$ quark mass
difference exceeds 4.1 MeV. Given all these clues, we expect that constituent
$m_d - m_u \simeq 4 {\rm~to~} 5.5 {\rm~MeV}$
will be fairly close to the truth.

Let us use the magnitude of $H_{tad}^3$ to come to some conclusions about
the magnitude of the $H_{JJ}$ piece for various baryons and thereby
estimate the strong interaction cutoff scale. The group-theoretical factors
ensuing from the Coleman-Glashow operator $H^3_{tad}$ and the
overall scale, estimated in ref. [10], provide the figures
\begin{equation}
(H^3_{tad})_{p-n} \simeq \frac{2}{3}(H^3_{tad})_{\Sigma^0-\Sigma^-}
\simeq \frac{1}{3}(H^3_{tad})_{\Sigma^+-\Sigma^-} \simeq
\frac{1}{2}(H^3_{tad})_{\Xi^0-\Xi^-} \simeq -2.5 {\rm~MeV}.
\end{equation}
Concentrating on the proton, we deduce that $(H_{JJ})_p \simeq 1.2$ MeV
in order to give the observed $n-p$ mass difference. Thus,
neglecting magnetic moment contributions to the fermion self-energy 
(which never exceed about 0.3 MeV) and using the standard QED result,
\begin{equation}
 (H_{JJ})_p \simeq \Sigma(m) = \frac{3\alpha}{2\pi}\left[
 \ln(\frac{\Lambda}{m_p}) + \frac{1}{4}\right] \simeq 1.2 {\rm~MeV},
\end{equation}
we require a strong interaction cut-off $\Lambda \simeq 1.05$ GeV.
This is a reasonable magnitude since it comes from vector-meson
dominated intermediate states and we will be adopting similar values
subsequently to estimate the photon exchange contributions to meson
masses. But in any case the picture looks rather good for baryons
when one also includes\cite{BCMT} the smaller magnetic contributions, as one
can see from Table 1.

\section{Pseudoscalar meson mass differences}


We ascribe the SU(2) differences to photon exchange and the $d-u$ quark
mass disparity in intermediate loops in order to see if we can arrive 
at the same sort of estimate as the group-theoretical tadpole method.
Turning first to the pions, it is readily established that the quark 
loop diagrams give (see Figure 1), at zero external momentum\cite{f4},
\begin{eqnarray}
(m^2_{\pi^+}-m^2_{\pi^0})_{qloops}&=& -8iN_c g^2\int \bar{d}^4p
                  \frac{p^2-m_um_d}{(p^2-m_u^2)(p^2-m_d^2)}\nonumber\\
& &+4iN_cg^2\int\bar{d}^4p\left[\frac{1}{p^2-m_u^2}+\frac{1}{p^2-m_d^2}
                  \right].
\end{eqnarray}
(We have dropped the isospin 1 $a_0(980)$ vacuum tadpole contribution,
because $a_0$ does not couple to pion pairs.) The result (8) equals
$(m_d-m_u)^2$ and may be neglected, in agreement with group-theoretical
symmetry arguments\cite{CG}. Not so the photon exchange
contribution\cite{Das} to the charged pion, which is quadratically
divergent in QED and, via dispersion relations, may be estimated to equal
\begin{equation}
(m^2_{\pi^\pm})_{JJ} \simeq \frac{\alpha}{2\pi}\left(
                     \Lambda^2+m_\pi^2[\ln(\Lambda^2/m_\pi^2)-1]\right).
\end{equation}
Since it is identified with $\Delta m_\pi^2 \simeq 1260 {\rm~MeV}^2$
experimentally, we require a cutoff $\Lambda\simeq 1.02$ GeV, rather close 
to the $p-n$ cutoff, used earlier. This is an encouraging sign.

Next we consider the kaons. Here, neither the quark bubble nor the $a_0$
tadpole graph is negligible and we need, as ever, the photon exchange
contribution,
\begin{equation}
 (m^2_{K^\pm})_{JJ} \simeq \frac{\alpha}{2\pi}\left(
                     \Lambda^2+m_K^2[\ln(\Lambda^2/m^2_K)-1]\right)
 \simeq 1420 {\rm~MeV}^2,
\end{equation}
again using a cutoff of about 1.05 GeV. The near agreement between (9) and 
(10) is compatible with Dashen's PCAC result\cite{RD}. 
The difference between quark loop contributions (see Figure 2) is
\begin{eqnarray}
(m^2_{K^+}-m^2_{K^0})_{qloops}&=&-8iN_cg^2\int\frac{\bar{d}^4p}{(p+k)^2-m_s^2}
\left[\frac{p.(p+k)-m_sm_u}{p^2-m_u^2}-\frac{p.(p+k)-m_sm_d}{p^2-m_d^2}
 \right]\nonumber\\
& & + \frac{8igg_{a_0KK}N_c}{m_{a_0}^2}\int \bar{d}^4p\left[
      \frac{m_u}{p^2-m_u^2} - \frac{m_d}{p^2-m_d^2}\right].
\end{eqnarray}
Using the log-divergent gap equation for the $s\bar{u}$ system\cite{DS},
namely $f_Kg=(m_s+\hat{m})/2$ or
\begin{equation}
 1 =-4iN_cg^2\int \frac{\bar{d}^4p}{(p^2-m_s^2)(p^2-\hat{m}^2)},
\end{equation}
and the regularization-insensitive identity \cite{DS},
$$\int\frac{\bar{d}^4p}{p^2-m^2}=\int\frac{m^2\,\bar{d}^4p}{(p^2-m^2)^2} +
       \frac{im^2}{16\pi^2}, $$
we are able to estimate the zero-momentum finite expression\cite{f2}:
\begin{eqnarray}
 (\Delta m_K^2)_{qloops} \equiv(m^2_{K^+}-m^2_{K^0})_{qloops}
 &\simeq &(m_u-m_d)\left[2(2\hat{m}-m_s)
  -3\hat{m}^2(\hat{m}-m_s)/2(\hat{m}^2+m_s^2) \right]
 \nonumber \\
 & & + 8g_{a_0KK}(m_u-m_d)\hat{m}^2/gm_{a_0}^2,
\end{eqnarray}
remembering that $g^2N_c = 4\pi^2$. Since the experimental magnitude is 
$(\Delta m_K^2)=(\Delta m_K^2)_{JJ}+(\Delta m_K^2)_{qloops}\simeq
-3960$ MeV$^2$, we require that $(\Delta m_K^2)_{qloops} \simeq -5380$ 
MeV$^2$. To make further progress we require some knowledge about
$g_{a_0KK}$. On the one hand we have the chiral estimate\cite{DS},
$g_{a_0KK} = (m_{a_0}^2-m_K^2)/2f_K \simeq$ 3140 MeV, and on the other
hand U(3) symmetry says that $g_{a_0KK}=g_{\sigma\pi\pi}/2=m_\sigma^2/2f_\pi
\simeq 2550$ MeV; probably the true value lies somewhere in between, say
$g_{a_0KK} \simeq 2700$ MeV with a possible error of 200 MeV.
Substituting this in (13), we are led to the value
$m_d-m_u \simeq 5380/1320 \simeq 4.1$ MeV, which is quite reasonable.

The same idea can be used to estimate the non-strange electromagnetic
transition amplitude,
$M_{\pi\eta_{ns}} = \langle \pi^0|H_{em}|\eta_{ns}\rangle$.
In this case we do not have to worry about photon exchange and the
surviving one-loop diagrams are given in Figure 3. Here one finds
$$M_{\pi\eta_{ns}} = -4ig^2N_c\int \bar{d}^4p\left[
     \frac{1}{p^2-m_u^2}-\frac{1}{p^2-m_d^2}\right] -
     \frac{8igg_{a_0\pi\eta_{ns}}}{m_{a_0}^2}\int \bar{d}^4p
     \left[\frac{m_u}{p^2-m_u^2} - \frac{m_d}{p^2-m_d^2}\right]. $$
We are on much firmer ground now if we claim that
$g_{a_0\pi\eta_{ns}}= m_\sigma^2/f_\pi \simeq 5140$ MeV, since this just
relies on U(2) symmetry. Using the gap equation, and $m_\sigma=2\hat{m}$,
we arrive at the clean result, $M_{\pi\eta_{ns}} = 2\hat{m}(m_u-m_d)
+ 16\hat{m}^3(m_u-m_d)/m_{a_0}^2 \simeq 1310 (m_u-m_d)$, and thereby
can predict the characteristic value $M_{\pi\eta_{ns}}\simeq -5240$
(MeV)$^2$, for $m_d - m_u \simeq 4$ MeV.

\section{Vector meson mass differences}


Next we turn to the vector mesons and the all-important coupling
between the $I=0,\,\,\omega$ and the $I=1,\,\,\rho$. The calculations
are even simpler in this case. Firstly we have the photon exchange
term, which comes out cleanly near the vector mass shell $k^2 = 
m_\rho m_\omega$ as
\begin{equation}
 (H_{JJ})_{\omega\rho}(k) = e^2k^2/(g_\rho g_\omega) \simeq
 640 {\rm~MeV}^2,
\end{equation}
since the leptonic rates give $g_\rho/e\simeq 16.6$ and $g_\omega/e\simeq 56.3$ via vector meson dominance.
Then we have the QED-like bubble polarization tensor term,
\begin{equation}
 \Pi_{\mu\nu}=(-k^2 g_{\mu\nu} +k_\mu k_\nu)\Pi(k^2,m_q^2)g_\rho^2/4
\end{equation}
where, in first approximation, we have used the U(2) symmetry coupling 
constants,
$ g_{\rho^0uu}=-g_{\rho^0dd}=g_{\omega uu}=g_{\omega dd}=g_\rho.$
The polarization function, being\cite{D}
\begin{equation}
 \Pi(k^2,m^2) = -8iN_c\int_0^1 d\alpha\int\frac{\alpha(1-\alpha)
                        \bar{d}^4p}{[p^2-m^2+k^2\alpha(1-\alpha)]^2},
\end{equation}
the difference between the $u$ and $d$ quark contributions is easily
found, because it is finite. At $k^2 = m_\rho^2 \simeq m_\omega^2$ one
obtains
\begin{eqnarray}
 \Pi(k^2,m_u^2)-\Pi(k^2,m_d^2) &=&-16iN_c(m_d^2-m_u^2)
  \int_0^1 \alpha(1-\alpha)\,d\alpha\int\frac{\bar{d}^4p}
      {[p^2+k^2\alpha(1-\alpha)]^3}\nonumber \\ 
   & = & \frac{N_c(m_d^2-m_u^2)}{2\pi^2k^2}.
\end{eqnarray}
Hence, via the inverse propagator $\Delta^{-1}_{\mu\nu}(k)=k_\mu k_\nu
 - k^2 g_{\mu\nu} + \Pi_{\mu\nu}(k) =-g_{\mu\nu}(k^2-m^2) + k_\mu
k_\nu$ terms, one sees that $-k^2\Pi(k^2)$ has the significance of
a squared mass. To this bubble contribution must be added the $a_0$ tadpole
contribution (see Figure 4 for the sum of all graphs), which equals
$$-\frac{4iN_cg_{a_0\omega\rho}}{m_{a_0}^2}\int \bar{d}^4p\left[
     \frac{m_u}{p^2-m_u^2} - \frac{m_d}{p^2-m_d^2}\right].$$
Let us invoke SU(4) spin-flavour symmetry and set $g_{a_0\pi\eta_{ns}}=
g_{a_0\rho\omega}$ in order to progress the evaluation;  in this manner
we estimate the $a_0$ tadpole contribution to equal that of the
$\pi-\eta_{ns}$ transition, namely $16\hat{m}^3(m_u-m_d)/m_{a_0}^2$.
Altogether, one deduces
\begin{equation}
 (H_{qloops})_{\omega\rho} = g_\rho^2N_c(m_u^2-m_d^2)/8\pi^2
              + 16\hat{m}^3(m_u-m_d)/m_{a_0}^2.
\end{equation}
The rhs can be estimated by using the experimental value $g_\rho =5.03$
\cite{DS1,Chan}; one finds that both terms in (18) contribute almost
equally, the result being
$(H_{qloops})_{\omega\rho} \simeq 1310(m_u-m_d) \simeq -5240$ MeV$^2$,
for $m_d-m_u\simeq 4$ MeV.
Adding it to the photon exchange contribution, we end up with
\begin{equation}
 (H_{em})_{\omega\rho}=(H_{JJ})_{\omega\rho}+(H_{qloops})_{\omega\rho}
  \simeq -4600 {\rm~MeV}^2.
\end{equation}
This agrees reasonably well with the magnitude $(H_{em})_{\omega\rho} 
\simeq -4520$ MeV$^2$, derived experimentally from the 
$\omega\rightarrow\rho^0\rightarrow 2\pi$ rate\cite{CB,CMS}.

We can extend these ideas to other mixings like $\rho J/\psi$,
$\rho \Upsilon$, but such calculations are sensitive to the amount
of admixture of nonstrange mesons in the heavy meson states. Indeed
the experimental rates for $J/\psi$ and $\Upsilon$ to two pions and
two kaons directly measure the admixtures---and they are very small.
Thus we unable to test properly our quark loop hypothesis in those 
cases. 

The calculations above confirm that one can view the Coleman-Glashow
tadpole piece as equivalent to a $d-u$ mass difference of about 4 MeV
in the context of a quark model. In this way we achieve a more fundamental
picture of the electromagnetic properties of hadrons, when we combine
the quark mass difference effect with standard photon emission/absorption.

To conclude this paper, we compare $\rho-\omega$ mixing with other methods
of estimation. Firstly there is the method based on the Coleman-Glashow
tadpole\cite{PS}:
\begin{equation}
(H_{tad}^3)_{\omega\rho} = -\langle 0|H_{tad}^3| a_0^0\rangle 
      g_{a_0\omega\rho^0}/m_{a_0}^2 \equiv -f_{a_0}m_{a_0}^2/f_\pi.
\end{equation}
For $f_{a_0} \simeq 0.5$ MeV, this gives $(H_{tad}^3)_{\omega\rho}
\simeq -5200$ MeV$^2$.  Secondly, one may use SU(3) symmetry to
connect this matrix element with the $K^*$ masses:
\begin{equation}
 (H_{tad}^3)_{\omega\rho} \simeq \Delta m^2_{K^*} - \Delta m_\rho^2
 \simeq -5130 {\rm~MeV}^2.
\end{equation}
Thirdly, one can apply fully fledged SU(6) symmetry to equate (21) with
\begin{equation}
 (H_{tad}^3)_{\omega\rho} \simeq \Delta m_K^2 -\Delta m_\pi^2
 \simeq -5220 {\rm~MeV}^2.
\end{equation}
In no case is there any striking discrepancy.  It might be possible also
to generalize the argument to heavier mesons like $D, D_s, D_c$ and $B$; this
would require strong faith in mass extrapolations and we have not been
brave enough to try that. In this connection it is worth recalling the result
of Weinberg and Lane\cite{LW}, based on phenomenological chiral Lagrangians,
which yield $m_d-m_u \sim 4.5$ MeV and which predicts $m_{D^+}-m_{D^0}
\simeq 6.7$ MeV.

As a parting note, observe that whether we use the Coleman-Glashow 
tadpole or regard it as the effect of an em quark loop, one is always
making contact with data in the (low-energy) s-channel. Alternatively,
Harari\cite{H} invoked crossing and duality to convert the view
into the (high-energy) t-channel. By studying superconvergent
relations, he identified the $\Delta I=1$ tadpole with a subtraction
constant in the t-channel, associated with the $\rho-$trajectory, and
thereby justified the Coleman-Glashow procedure.

\acknowledgments
We thank the Australian Research Council for providing a grant which 
enabled this collaboration to take place. MDS appreciates the
hospitality of the University of Tasmania.

\begin{table}[h]
\caption{SU(2) mass splittings for octet baryons, in MeV.}
\begin{tabular}{|l|l|l|l|l|}
Baryons & $H_{JJ}$ & $H^3_{tad}$ & Total & Data\\
\hline
$m_p-m_n$ & 1.2 &-2.5 & -1.3 & -1.29\\
$m_{\Sigma^0}-m_{\Sigma^-}$ & -1.0 & -3.8 & -4.8 & $-4.9\pm 0.1$\\
$m_{\Sigma^+}-m_{\Sigma^-}$ & -0.3 & -7.5 & -7.8 & $-8.1\pm 0.1$\\
$m_{\Xi^0}-m_{\Xi^-}$ & -1.1 & -5.0 & -6.1 & $-6.4\pm 0.6$
\end{tabular}

\end{table}

\newpage

\noindent Figure 1. Quark loop plus photon exchange contributions to
$m_{\pi^+}-m_{\pi^-}$.

\noindent Figure 2. Quark loop contributions to the $\pi-\eta_{ns}$
transition element.

\noindent Figure 3. Quark loop plus photon exchange contributions to
$m_{K^0}-m_{K^+}$.

\noindent Figure 4. Quark loop contrbutions to the $\rho-\omega$
transition element.

\setlength{\unitlength}{1cm}

\newpage

\begin{figure}
\begin{picture}(14,20)
\put (0.5,0.0){\epsfxsize=13cm \epsfbox{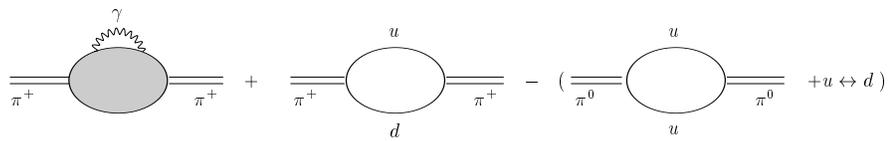}}
\end{picture}
\caption{R Delbourgo}
\label{fig:f1}
\end{figure}

\newpage

\begin{figure}
\begin{picture}(14,20)
\put (0.5,0.0){\epsfxsize=13cm \epsfbox{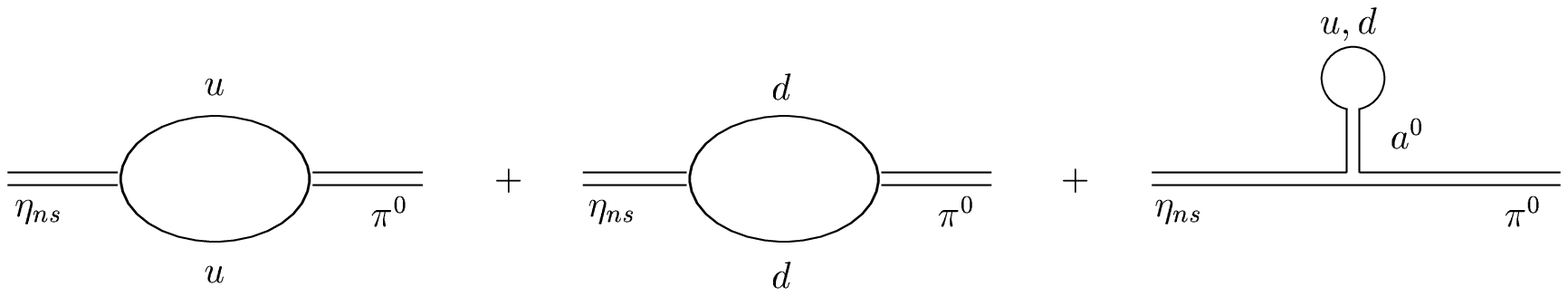}}
\end{picture}
\caption{R Delbourgo}
\label{fig:f2}
\end{figure}

\newpage

\begin{figure}
\begin{picture}(14,20)
\put (0.5,0.0){\epsfxsize=13cm \epsfbox{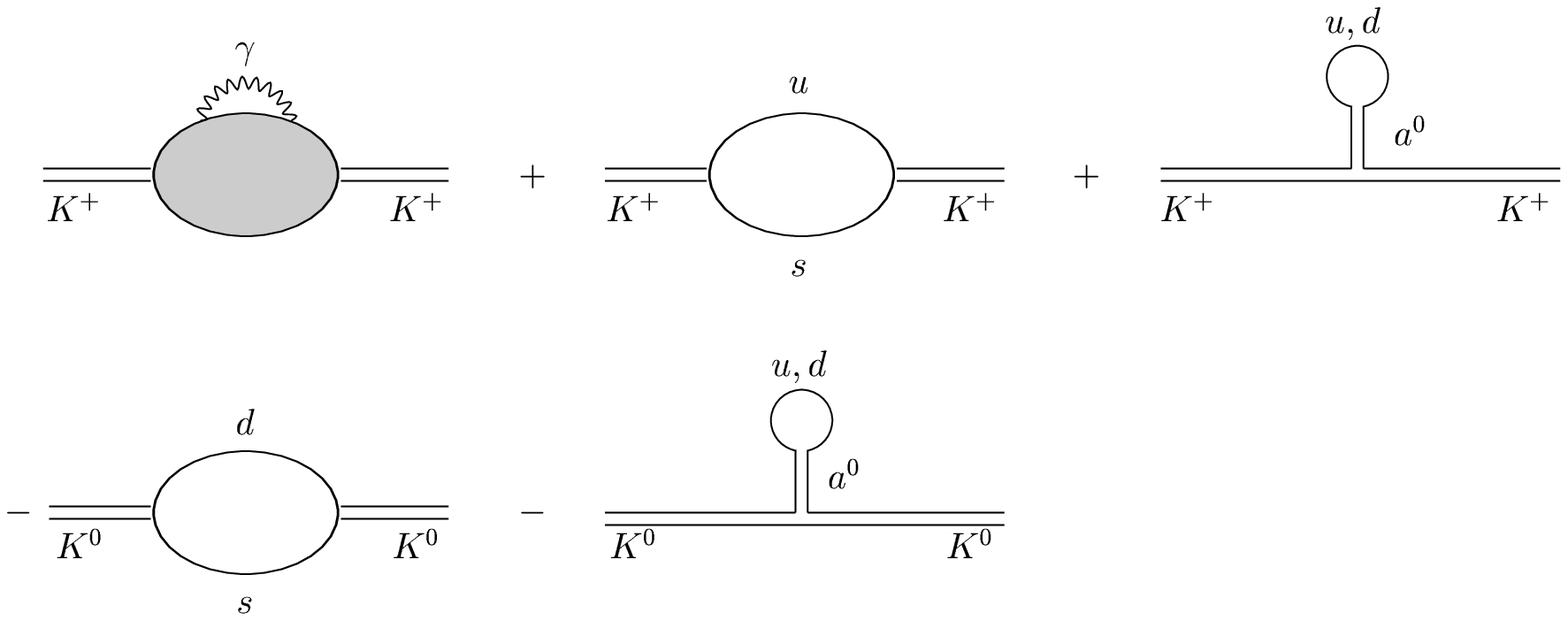}}
\end{picture}
\caption{R Delbourgo}
\label{fig:f3}
\end{figure}

\newpage

\begin{figure}
\begin{picture}(14,20)
\put (0.5,0.0){\epsfxsize=13cm \epsfbox{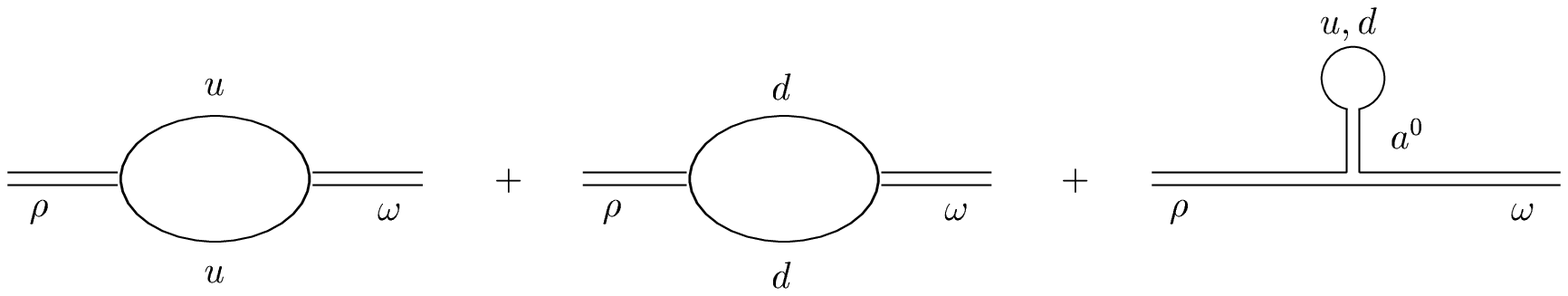}}
\end{picture}
\caption{R Delbourgo}
\label{fig:f4}
\end{figure}


\begin{references}
\bibitem{f2} It is a surprising fact that meson interactions are well
determined by quark loops at zero external momentum; for the lighter
mesons like $u,d,s$, this is no great deal as the extrapolation from
the mass shell to zero momentum is small. The amazing thing is that
trlinear meson couplings for the heavier quark $c$ composites are also
well predicted in this way. It is as though there is an underlying
renormalization mass scale which is {\em much} larger than $\Lambda_{QCD}$
which renders the continuation innocuous.
\end{references}
\end{document}